\documentclass[aps,prl,showpacs,twocolumn,superscriptaddress]{revtex4}
\usepackage{amsmath}
\usepackage{amssymb}
\usepackage{epsfig}
\usepackage{color}
\usepackage{graphicx,amsmath}

\begin{document}
\title{Thermodynamic Properties of Kagome Lattice in $\rm ZnCu_3(OH)_6Cl_2$
Herbertsmithite}
\author{V.R. Shaginyan}\email{vrshag@thd.pnpi.spb.ru}
\affiliation{Petersburg Nuclear Physics Institute,  Gatchina,
188300, Russia}\affiliation{Clark Atlanta University, Atlanta, GA
30314, USA} \author{A.Z. Msezane}\affiliation{Clark Atlanta
University, Atlanta, GA 30314, USA}\author{K.G.
Popov}\affiliation{Komi Science Center, Ural Division, RAS,
Syktyvkar, 167982, Russia}

\begin{abstract}
Strongly correlated Fermi systems are among the most intriguing and
fundamental systems in physics, whose realization in some compounds
is still to be discovered. We show that herbertsmithite $\rm
ZnCu_3(OH)_6Cl_2$ can be viewed as a strongly correlated Fermi
system whose low temperature thermodynamic in magnetic fields is
defined by a quantum critical spin liquid. Our calculations of its
thermodynamic properties are in good agreement with recent
experimental facts and allow us to reveal their scaling behavior
which strongly resembles that observed in HF metals and 2D $\rm ^3
He$.
\end{abstract}


\maketitle

An explanation of the rich behavior of strongly correlated Fermi
systems still continues to be among the main problems of the
condensed matter physics. One of the most interesting and puzzling
issues in the research of strongly correlated Fermi systems is the
non-Fermi liquid (NFL) behavior detected in their thermodynamic
properties. Under the application of external fields, e.g. magnetic
field $B$, the system can be driven to a Landau Fermi liquid
behavior (LFL). Such a behavior was observed in quite different
objects such as heavy-fermion (HF) metals \cite{loh,pr} and
two-dimensional $\rm ^3He$ \cite{3he,3he1,prl,pr}

Recently the herbertsmithite $\rm ZnCu_3(OH)_6Cl_2$ has been
exposed as a $S=1/2$ kagome antiferromagnet \cite{herb} and new
experimental investigations have revealed its unusual behavior
\cite{herb1,herb2,herb3}. Because of the electrostatic environment,
$\rm Cu^{2+}$ is expected to occupy the distorted octahedral kagome
sites. Magnetic kagome planes $\rm Cu^{2+}$ $S=1/2$ are separated
by nonmagnetic $\rm Zn^{2+}$ layers. Observations have found no
evidence of long range magnetic order or spin freezing down to
temperature of 50 mK, indicating that $\rm ZnCu_3(OH)_6Cl_2$ is the
best model found of quantum kagome lattice
\cite{mil,herb1,herb2,herb3}. The specific heat $C$, arising from
the $\rm Cu$ spin system,  at $T<1$ K appears to be governed by a
power law with an exponent which is less than or equal to 1. At the
lowest explored temperature, namely over the temperature range
$106<T<400$ mK, $C$ follows a linear law temperature dependence,
$C\propto T$, and for temperatures of a few Kelvin and higher, the
specific heat becomes $C(T)\propto T^3$ and is dominated by the
lattice contribution \cite{herb1,herb2,herb3}. At low temperatures
$T\leq 1$, the strong magnetic field dependence of the specific
heat $C$ suggests that $C$ is predominately magnetic in origin
\cite{herb1,herb2,herb3}. There are a number of papers suggesting
that the $S=1/2$ model on the kagome lattice can be viewed as the
gapless critical spin liquid
\cite{herb1,herb2,herb3,sl,sl1,sl2,sl3}. These facts allow us to
test both the NFL and LFL behavior of $\rm ZnCu_3(OH)_6Cl_2$ and to
show that a Fermi quantum spin liquid formed in the herbertsmithite
determines its low temperature thermodynamic properties.

\begin{figure} [! ht]
\begin{center}
\includegraphics [width=0.47\textwidth]{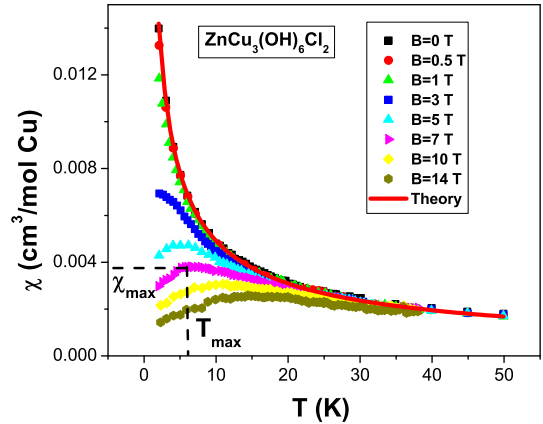}
\end{center}
\caption{Temperature dependence of the magnetic susceptibility
$\chi$ at different magnetic fields for $\rm ZnCu_3(OH)_6Cl_2$
\cite{herb3}. The illustrative values of $\chi_{\rm max}$ and
$T_{\rm max}$ at $B=7$ T are also shown. Our calculations made at
$B=0$ are depicted by the solid curve representing $\chi(T)\propto
T^{-\alpha}$ with $\alpha=2/3$.}\label{fig1}
\end{figure}

Contrary to the $C\propto T$ behavior \cite{herb1,herb2}, the
observed spin liquids contribute a $T^2$ specific heat which in the
model is not sensitive to an applied magnetic field \cite{sl2,sl3}.
Moreover the magnetic susceptibility $\chi(T)$ of $\rm
ZnCu_3(OH)_6Cl_2$ shown in Fig. \ref{fig1} displays an unusual
behavior \cite{herb3}. At $B\geq 3$ T, $\chi(T)$ has a maximum
$\chi_{\rm max}(T)$ at some temperature $T_{\rm max}(B)$. The
maximum $\chi_{\rm max}(T)$ decreases as magnetic field $B$ grows,
while $T_{\rm max}(B)$ shifts to higher $T$ reaching $15$ K at
$B=14$ T. At $B\leq 1$ T as seen from Fig. \ref{fig1},
$\chi(T)\propto T^{-\alpha}$ with $\alpha=2/3$. The calculated
exponent \cite{pr,ckz} is in good agreement with the experimental
value $\alpha=2/3\simeq 0.66$ \cite{herb3}.  The observed behavior
of $\chi$ strongly resembles that in HF metals and is associated
with their proximity to a quantum critical point (QCP)
\cite{pr,ckz,khodb}.
As a result, we safely assume that a quantum deconfined Fermi
critical spin liquid with essentially gapless excitations formed by
neutral fermions is realized in $\rm ZnCu_3(OH)_6Cl_2$ and located
very near QCP \cite{herb1}. Thus, $\rm ZnCu_3(OH)_6Cl_2$ turns out
to be located at its QCP without tuning this substance to QCP using
a control parameter such as magnetic field, pressure, or chemical
composition. This observation is in sharp contrast to the common
practice applied to tune HF metals to their QCP's. A simple kagome
lattice may have a dispersionless topologically protected branch of
the spectrum with zero excitation energy that is the flat band
\cite{green,vol}. In that case a fermion condensation quantum phase
transition (FCQPT) \cite{pr} can be considered as QCP of the $\rm
ZnCu_3(OH)_6Cl_2$ quantum spin liquid.

In this letter we uncover the quantum spin liquid phase and its QCP
in the herbertsmithite $\rm ZnCu_3(OH)_6Cl_2$ and explain its low
temperature thermodynamic in magnetic fields. For the first time we
calculate the susceptibility $\chi$, magnetization $M$ and specific
heat $C$ as functions of temperature $T$ versus magnetic field $B$.
Our calculations are in good agreement with the experimental facts
and allow us to reveal their scaling behavior which strongly
resembles that observed in HF metals and 2D $\rm ^3 He$.

To study the low temperature thermodynamic and scaling behavior, we
use the model of homogeneous heavy-fermion liquid \cite{pr}. This
model permits to avoid complications associated with the
crystalline anisotropy of solids. We propose that the critical
quantum spin liquid is composed of fermions. These fermions with
zero charge and spin $\sigma=1/2$ occupy the corresponding Fermi
sphere with the Fermi momentum $p_F$. The ground state energy
$E(n)$ is given by the Landau functional depending on the
quasiparticle distribution function $n_\sigma({\bf p})$, where
${\bf p}$ is the momentum. Near FCQPT point, the effective mass
$M^*$ is governed by the Landau equation \cite{land,pr}
\begin{eqnarray}
\nonumber
&&\frac{1}{M^*(T,B)}=\frac{1}{M^*}+\frac{1}{p_F^2}\sum_{\sigma_1}
\int\frac{{\bf p}_F{\bf p_1}}{p_F}\\
&\times&F_{\sigma,\sigma_1}({\bf p_F},{\bf
p}_1)\frac{\partial\delta n_{\sigma_1}({\bf p}_1,T,B)}
{\partial{p}_1}\frac{d{\bf p}_1}{(2\pi) ^3}. \label{HC3}
\end{eqnarray}
Here we have rewritten the quasiparticle distribution function as
$n_{\sigma}({\bf p},T,B) \equiv n_{\sigma}({\bf p},T=0,B=0)+\delta
n_{\sigma}({\bf p},T,B)$. The Landau amplitude $F$ is completely
defined by the fact that the system has to be at QCP of FCQPT
\cite{pr,ckz,epl}. The sole role of the Landau amplitude is to
bring the system to FCQPT point, where Fermi surface alters its
topology so that the effective mass acquires temperature and field
dependence \cite{pr,ckz,khodb}. At this point, the term $1/M^*$
vanishes and Eq. \eqref{HC3} becomes homogeneous. It can then be
solved analytically \cite{pr,ckz}. At $B=0$, the effective mass
strongly depends on $T$ demonstrating the NFL behavior
\cite{pr,ckz}
\begin{equation}
M^*(T)\simeq a_TT^{-2/3}.\label{MTT}
\end{equation}
At finite $T$, the application of magnetic field $B$ drives the
system the to LFL region with
\begin{equation}
M^*(B)\simeq a_BB^{-2/3}.\label{MBB}
\end{equation}

At finite $B$ and $T$ near FCQPT, the solutions of Eq. \eqref{HC3}
$M^*(B,T)$ can be well approximated by a simple universal
interpolating function. The interpolation occurs between the LFL
($M^*(T)\propto const)$ and NFL ($M^*(T)\propto T^{-2/3}$) regions
\cite{pr,ckz}. It is convenient to introduce the normalized
effective mass $M^*_N$ and the normalized temperature $T_N$
dividing the effective mass $M^*$ by its maximal values, $M^*_M$,
and temperature $T$ by $T_{\rm max}$ at which the maximum occurs.
Equation \eqref{HC3} allows us to calculate the thermodynamic
properties for the normalized susceptibility $\chi_N=\chi/\chi_{\rm
max}=M^*_N$. Since $C/T\propto M^*$, the normalized
$(C/T)_N=\chi_N=M^*_N$. We note that our calculations of $M^*_N$
based on Eq. \eqref{HC3} do not contain any free fitting
parameters. The normalized effective mass $M^*_N=M^*/M^*_M$ as a
function of the normalized temperature $y=T_N=T/T_{\rm max}$ is
given by \cite{pr,ckz}
\begin{equation}M^*_N(y)\approx c_0\frac{1+c_1y^2}{1+c_2y^{8/3}}.
\label{UN2}
\end{equation}
Here $c_0=(1+c_2)/(1+c_1)$, $c_1$ and $c_2$ are fitting parameters,
approximating the Landau amplitude. Magnetic field $B$ enters Eq.
\eqref{HC3} only in the combination $\mu_BB/k_BT$, making
$k_BT_{\rm max}\simeq \mu_BB$ where $k_B$ is the Boltzmann constant
and $\mu_b$ is the Bohr magneton \cite{ckz,pr}. Thus, in the
presence of magnetic fields the variable $y$ becomes
\begin{equation}\label{YTB}
y=T/T_{\rm max}\simeq k_BT/\mu_BB.
\end{equation}
The variables $T$ and $B$ enter Eq. \eqref{YTB} symmetrically;
therefore Eq. \eqref{UN2} is valid for $y=\mu_BB/k_BT$. In what
follows we use Eq. \eqref{UN2} to clarify our calculations based on
Eq. \eqref{HC3}. It follows directly from Eqs. \eqref{MBB},
\eqref{UN2} and \eqref{YTB} that $\chi(k_BT/\mu_BB)T^{2/3}\propto
y^{2/3}M^*_N(y)$. Since the magnetization $M(B,T)=\int
\chi(B,T)dB$, we obtain that $M(B,T)T^{-1/3}$ depends on the only
variable $y$. These observations confirm the scaling behavior of
both $\chi T^{0.66}$ and $M T^{-0.34}$ that is experimentally
established in \cite{herb3}.

\begin{figure}[!ht]
\begin{center}
\includegraphics [width=0.40\textwidth]{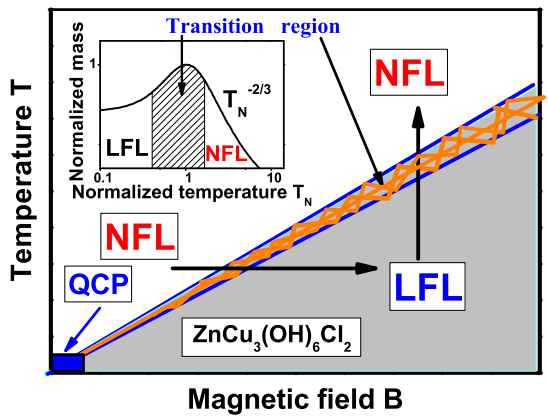}
\end{center}
\caption{Schematic phase diagram of $\rm ZnCu_3(OH)_6Cl_2$. The
vertical and horizontal arrows show LFL-NFL and NFL-LFL transitions
at fixed $B$ and $T$ respectively. Inset shows a schematic plot of
the normalized effective mass versus the normalized temperature.
Transition region, where $M^*_N$ reaches its maximum at
$T_N=T/T_{\rm max}=1$, is shown by the arrows and hatched area in
both the main panel and in the inset.}\label{fig0}
\end{figure}
We are now in a position to construct the schematic phase diagram
of $\rm ZnCu_3(OH)_6Cl_2$. The phase diagram is reported in Fig.
\ref{fig0}. At $T=0$ and $B=0$ the system is located at QCP of
FCQPT without tuning. Both magnetic field $B$ and temperature $T$
play the role of the control parameters, shifting the system from
its QCP and driving it from the NFL to LFL regions as shown by the
vertical and horizontal arrows. At fixed temperatures the increase
of $B$ drives the system along the horizontal arrow from the NFL
region to LFL one. On the contrary, at fixed magnetic field and
increasing temperatures the system transits along the vertical
arrow from the LFL region to NFL one. The inset to Fig. \ref{fig0}
demonstrates the behavior of normalized effective mass $M^*_N$
versus normalized temperature $T_N$ following from Eq. \eqref{UN2}.
It is seen that the temperature region $T_N\sim 1$ represents a
transition region between the LFL behavior with almost constant
effective mass and the NFL one, having the $T^{-2/3}$ dependence.
It is seen from Eqs. \eqref{UN2} and \eqref{YTB} and Fig.
\ref{fig0} that the width of the transition region $T_w\propto
T\propto B$.

\begin{figure} [! ht]
\begin{center}
\includegraphics [width=0.47\textwidth]{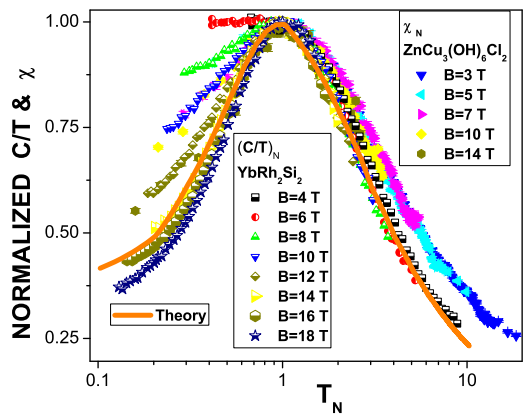}
\end{center}
\caption{Normalized susceptibility $\chi_N=\chi/\chi_{\rm
max}=M^*_N$ versus normalized temperature $T_N$. $\chi_N$ is
extracted from the measurements of the magnetic susceptibility
$\chi$ in magnetic fields $B$ \cite{herb3} shown in Fig.
\ref{fig1}. Normalized specific heat $(C/T)_N=M^*_N$ is extracted
from the measurements of $C/T$ on $\rm YbRh_2Si_2$ in magnetic
fields $B$ \cite{steg1}. The corresponding fields $B$ are listed in
the legends. Our calculations made at field $B$ completely
polarizing the quasiparticle band are depicted by the solid curve
tracing the scaling behavior of $M^*_N$.}\label{fig2}
\end{figure}
The experimental data on measurements of $\chi_N$ \cite{herb3},
$(C/T)_N=M^*_N$ \cite{steg1} and our calculations of $M^*_N$ at
fixed magnetic field $B$ that completely polarizes the
quasiparticle band are shown respectively by the geometrical
figures and solid curve in Fig. \ref{fig2}. It is clearly seen that
the data collected on both $\rm ZnCu_3(OH)_6Cl_2$ and $\rm
YbRh_2Si_2$ collapse into the same curve, obeying the scaling
behavior. Consistent with the phase diagram displayed in Fig.
\ref{fig0}, at growing temperatures ($y\simeq 1$) the LFL behavior
first converts into the transition one and then disrupts into the
NFL behavior. This demonstrates that the spin liquid of $\rm
ZnCu_3(OH)_6Cl_2$ is close to QCP and behaves as the HF liquid of
$\rm YbRh_2Si_2$. It is seen, that the low-temperature ends ($T_N
\leq 0.5$) of the curves do not merge and their values decrease as
$B$ grows representing the full spin polarization of the HF band at
the highest reached magnetic fields \cite{epl}. Indeed, at low
$T_N$, $\chi_N$ at $B=14$ T is close to $(C/T)_N$ at $B=18$ T,
while our calculations shown by the solid curve is close to both
functions.

\begin{figure} [! ht]
\begin{center}
\includegraphics [width=0.47\textwidth]{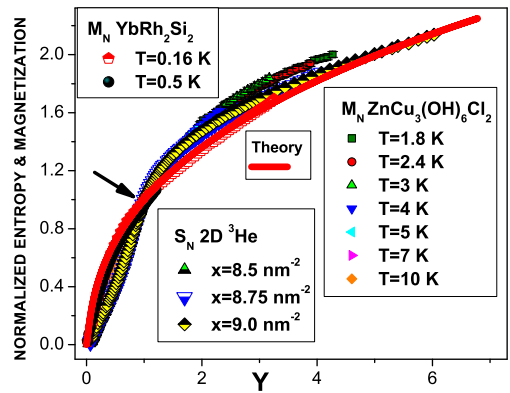}
\end{center}
\caption{Normalized magnetization $M_N(y)$ collected on
measurements on $\rm ZnCu_3(OH)_6Cl_2$ \cite{herb3} and $\rm{
YbRh_2Si_2}$ \cite{steg} at different temperatures shown in the
corresponding legends. Shown by the arrow a kink is clearly seen at
$y\simeq 1$. The normalized entropy $S_N(y)$ is extracted from
measurements on 2D $\rm ^3He$ \cite{3he} at different densities
${\rm x}$ shown in the legend. The solid curve represents our
calculations of the normalized magnetization.}\label{fig3}
\end{figure}
Both the normalized magnetization $M_N(y)=M(B/B_k)/M(B_k)$,
extracted from measurements of the magnetization $M(B)$
\cite{herb3}  depicted by the geometrical figures, and calculated
$M_N(y)$ shown by the solid line are reported in Fig. \ref{fig3}.
Here, $T_k$ is the temperature at which the magnetization
demonstrates the kink, while the system enters the transition
region shown in Fig. \ref{fig0} \cite{pr}. The normalized entropy
$S_N(y)=S(T/T_{inf})/S(T_{inf})$ is obtained from measurements of
the entropy $S$ on 2D $\rm ^3He$ \cite{3he}. Here $T_{inf}$ is the
temperature at which the system enters the transition region and
$S$ possesses its inflection point, clearly seen in the data (see
the Supporting Online Material for \cite{3he}, Fig. S8, A). It is
seen from Eq. \eqref{UN2} and from the inset to Fig. \ref{fig0},
that at $y<1$, $S_N=M_N\propto y$, and at $y>1$, $S_N=M_N\propto
y^{1/3}$. This behavior produces the kink and makes the scaled data
merge into a single curve in terms of the variable $y$ \cite{pr}.
Our calculations are in good agreement with the measurements.
\begin{figure} [! ht]
\begin{center}
\includegraphics [width=0.47\textwidth]{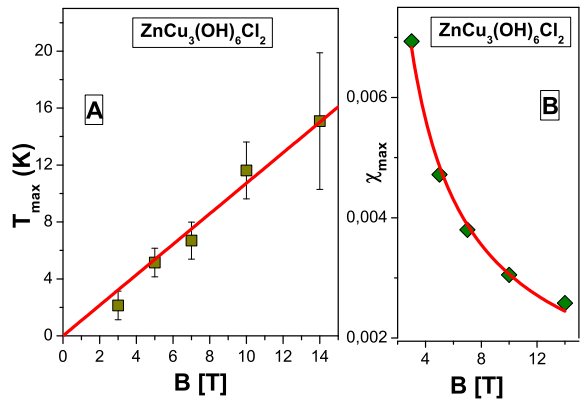}
\end{center}
\caption{Panel {\bf A}: The temperatures $T_{\rm max}(B)$ at which
the maxima of $\chi$ (see Fig. \ref{fig1}) are located. The solid
line represents the function $T_{\rm max}\propto aB$, $a$ is a
fitting parameter, see Eq. \eqref{YTB}. Panel {\bf B}: The maxima
$\chi_{\rm max}$ of the function $\chi(T)$ versus magnetic field
$B$ (see Fig. \ref{fig1}). The solid curve is approximated by
$\chi_{\rm max}(B)=dB^{-2/3}$, see Eq. \eqref{MBB}, $d$ is a
fitting parameter.}\label{fig4}
\end{figure}

\begin{figure} [! ht]
\begin{center}
\includegraphics [width=0.47\textwidth]{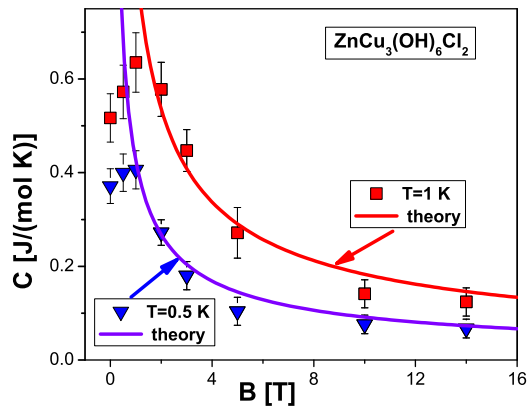}
\end{center}
\caption{The specific heat $C(B,T)$ versus magnetic field $B$
measured on $\rm ZnCu_3(OH)_6Cl_2$ at two different temperatures
$T$ \cite{herb1} listed in the legends is shown by the triangles
and squares. Our calculations are depicted by the solid curves
tracing the LFL behavior of $C(B,T)=a_1B^{-2/3}T$, see Eq.
\eqref{MBB}, with $a_1$ being the fitting parameter.}\label{fig6}
\end{figure}
In Fig. \ref{fig4}, panel {\bf A}, the solid squares denote
temperatures $T_{\rm max}(B)$ at which the maxima of $\chi(T)$
occur versus magnetic field $B$. In the panel {\bf B}, the
corresponding values of the maxima $\chi_{\rm max}(B)$ are shown by
the solid diamonds versus $B$. It is seen that the agreement
between the theory and experiment is good in the entire magnetic
field domain. Our calculations of the specific heat $C(B,T)$ are
shown in Fig. \ref{fig6}. For $T$ of a few Kelvin and higher, the
lattice contribution to the specific heat is the most significant
contribution. However, this contribution diminishes at low
temperatures, and at $T\leq1$ K $C$ is predominately formed by the
spin liquid \cite{herb1}. It is seen from Fig. \ref{fig6}, that in
the LFL region at  $k_BT\lesssim \mu_BB$, $C(B,T)\propto
M^*T\propto B^{-2/3}T$, and field $B$ completely defines the
$M^*(B)$ behavior given by Eq. \eqref{MBB}. Clearly, our
calculations are in good agreement with the measurements when the
system demonstrates the LFL behavior. Indeed, at $T=1$ K the system
exhibits the LFL behavior at $B\geq 2$ T, while at $T=0.5$ K the
LFL behavior is observed even at lower values of $B$, namely $B\geq
1$ T.

In summary, we have shown that the kagome lattice of $\rm
ZnCu_3(OH)_6Cl_2$ can be viewed as a strongly correlated Fermi
system whose thermodynamic is defined by the quantum critical spin
liquid located at FCQPT. Our calculations of the thermodynamic
properties are in good agreement with the experimental facts and
their scaling behavior coincides with that observed in HF metals and
2D $\rm ^3 He$. We have also demonstrated that $\rm
ZnCu_3(OH)_6Cl_2$ exhibits the LFL, NFL and the transition behavior
as HF metals and 2D $\rm ^3 He$ do.

We grateful to  V.A. Khodel and V.A. Stephanovich for valuable
discussions. This work was supported by U.S. DOE, Division of
Chemical Sciences, Office of Basic Energy Sciences, Office of
Energy Research, AFOSR and the RFBR \# 09-02-00056.


\begin{thebibliography}{99}

\bibitem{loh} H.v. L\"ohneysen {\it et al.,}
\rmp {\bf 79}, 1015 (2007).

\bibitem{pr} V.R. Shaginyan, M.Ya. Amusia, A.Z. Msezane, and
K.G. Popov,  Phys. Rep. {\bf 492}, 31 (2010).

\bibitem{3he} M. Neumann, J. Ny\'{e}ki, and J. Saunders, Science {\bf
317}, 1356 (2007).

\bibitem{3he1} R. Masutomi {\it et al.,} \prl {\bf 92}, 025301 (2004).

\bibitem{prl}  V.R. Shaginyan {\it et al.,} \prl {\bf 100}, 096406 (2008).

\bibitem{herb} M.P. Shores {\it et al.,}
J. Am. Chem. Soc. {\bf 127}, 13462 (2005).

\bibitem{herb1} J.S. Helton {\it et al.,} \prl {\bf 98}, 107204 (2007).

\bibitem{herb2} M.A. de Vries {\it et al.,} \prl {\bf 100}, 157205 (2008).

\bibitem{herb3} J.S. Helton {\it et al.,} \prl {\bf 104}, 147201 (2010).

\bibitem{mil} F. Mila, \prl {\bf 81}, 2356 (1998).

\bibitem{green} D. Green, L. Santos, and C. Chamon, \prb {\bf 82}, 075104
(2010).

\bibitem{vol} G.E. Volovik, arXiv:1012.0905v3.

\bibitem{sl} S.S. Lee and P.A. Lee, \prl {\bf 95}, 036403 (2005).

\bibitem{sl1} O.I. Motrunich, \prb {\bf 72}, 045105 (2005).

\bibitem{sl2} Y. Ran {\it et al}., \prl {\bf 98},
117205 (2007).

\bibitem{sl3} S. Ryu {\it et al}., \prb {\bf 75}, 184406 (2007).

\bibitem{ckz} J.W. Clark, V.A. Khodel, and M.V. Zverev
\prb {\bf 71}, 012401 (2005).

\bibitem{khodb} V.A. Khodel, J.W. Clark, and M.V. Zverev,
\prb {\bf 78}, 075120 (2008).

\bibitem{land} L.D. Landau, Sov. Phys. JETP {\bf 3}, 920 (1956).

\bibitem{steg1} P. Gegenwart {\it et al}., New J. Phys. {\bf 8}, 171 (2006).

\bibitem{epl} V. R. Shaginyan {\it et al}., Europhys.
Lett. {\bf 93}, 17008 (2011).

\bibitem{steg}  P. Gegenwart {\it et al}., Science {\bf 315}, 969 (2007).

\end{thebibliography}
\end{document}